\documentclass[twocolumn,usenatbib]{mn2e}
\usepackage{amsmath}
\usepackage{epsfig}
\usepackage{graphicx}
\usepackage{amssymb}
\begin{document}

\title{An Infinite Family of Self-consistent Models for Axisymmetric Flat Galaxies}

\author[J. F. Pedraza, J. Ramos-Caro and G. A. Gonz\'alez]
{Juan F. Pedraza\thanks{E-mail: jfpa080@gmail.com}, Javier
Ramos-Caro\thanks{E-mail: javiramos1976@gmail.com} and
Guillermo A. Gonz\'alez\thanks{E-mail: gonzalez@gag-girg-uis.net} \\
Escuela de F\'isica,  Universidad Industrial de Santander, A. A. 678,
Bucaramanga, Colombia}

\maketitle

\begin{abstract}

We present the formulation of a new infinite family of self-consistent stellar
models, designed to describe axisymmetric flat galaxies. The corresponding
density-potential pair is obtained as a superposition of members belonging to
the generalized Kalnajs family, by imposing the condition that the density can
be expressed as a regular function of the gravitational potential, in order to
derive analytically the corresponding equilibrium distribution functions (DF).
The resulting models are characterized by a well-behaved surface density, as in
the case of generalized Kalnajs discs. Then, we present a study of the
kinematical behavior which reveals, in some particular cases, a very
satisfactory behavior of the rotational curves (without the assumption of a dark
matter halo). We also analyze the equatorial orbit's stability and Poincar\'e
surfaces of section are performed for the 3-dimensional orbits. Finally, we
obtain the corresponding equilibrium DFs, using the approaches introduced by
\cite{kal} and \cite{dej}.

\end{abstract}


\section{Introduction}

The obtention of density-potential pairs (PDP) corresponding to idealized thin
discs is a problem of great astrophysical relevance, motivated by the fact that
the main part of the mass in many galaxies is concentrated in an stellar flat
distribution, usually assumed as axisymmetric (\cite{BT}). Once the
potential-density pair (PDP) is formulated as a model for a galaxy, usually the
next step is to find the corresponding distribution function (DF). This is one
of the fundamental quantities in galactic dynamics specifying the distribution
of the stars in the phase-space of positions and velocities. Although the DF can
generally not be measured directly, there are some observationally accesible
quantities that are closed related to the DF: the projected density and the
light-of-sight velocity, provided by photometric and kinematic observations, are
examples of DF's moments. Thus, the formulation of a PDP with its corresponding
equilibrium DFs establish a self-consistent stellar model that can be
corroborated by astronomical observations.

Now then, there is a variety potential-density pairs for such flat stellar
models, e.g. \cite{WM}; \cite{KUZ}; \cite{Schmith}; \cite{T1,T2}; \cite{BB};
\cite{KAL}; \cite{GR}. In particular, \cite{T1,T2} formulated a generalized
family of models whose first member is the one introduced by \cite{KUZ}. This
family represents a set of discs of infinite extension, derived by solving the
Laplace equation  in cylindrical coordinates subject to appropriated boundary
conditions on the discs and at infinity.

Analogously, \cite{GR} obtained a family of finite thin discs (generalized
Kalnajs discs) whose first member corresponds precisely to the well-known model
derived by \cite{KAL}. Such family was derived by using the Hunter's method
(\citeyear{HUN1}), which is based in the obtention of solutions of Laplace
equation in terms of oblate spheroidal coordinates, by imposing some appropriate
conditions on the surface density. So, by requiring that the surface density
behaves as a monotonously decreasing function of the radius, with a maximum at
the center of the disk and vanishing at the edge, detailed expressions for the
gravitational potential and the rotational velocity were obtained as series of
elementary functions. Also, some two-integral distribution functions for the
first four members of this family were recently obtained by \cite{PRG}. Now
then, as the generalized Kalnajs models correspond to discs of finite extension,
they can be considered as more realistic descriptions of flat galaxies than the
Toomre's family.

In the present paper we formulate a new infinite set of finite thin discs,
obtained by superposing the members of the generalized Kalnajs family in such a
way that the resulting density surface can be expressed as a well behaved
function of the gravitational potential. As it was pointed out by some authors,
this is a fundamental requirement for the searching of equilibrium distribution
functions (DF) describing such axisymmetric systems (see for example,
\cite{fricke}, \cite{Hunter}, \cite{jiang}). Thus, the new family formulated
here has the advantage of easily providing the corresponding two-integral DFs.

Furthermore, the models have two additional advantages. In one hand, the mass
surface density is well-behaved, as in the case of generalized Kalnajs discs,
having a maximum at the center and vanishing at the edge. Moreover, the mass
distribution of the higher members of the family is more concentrated  at the
center. On the other hand, the rotation curves are better behaved than in the
Kalnajs discs. We found that, in some cases, the circular velocity increases
from a value of zero at the center of the disc, then reaches a maximum at some
critical radius and, after that, remains approximately constant. As it is known,
such behavior has been observed in  many disclike  galaxies.

Now, apart from the circular velocity, there are two important quantities
concerning to the interior kinematics of the models: the epicyclic and vertical
frequencies, which describe the stability against radial and vertical
perturbations of particles in quasi-circular orbits. We found that the models
formulated here are radially stable whereas vertically unstable, which is a
characteristic inherited from the generalized Kalnajs family (\cite{ramos}).
When we deal with three dimensional orbits, there are also a common feature
between the new models and the Kalnajs family: the phase space structure of
disc-crossing orbits, that can be viewed through the Poincar\'e surfaces of
section, is composed by shape-of-ring KAM curves and prominent chaotic zones.
However, there are certain situations in which the chaoticity disappears and the
3-dimensional motion of test particles is completely regular. In such cases, one
can suggest the existence of a third integral of motion, as in the case of
St\"{a}ckel and Kuzmin potentials.

Finally, in order to formulate the new family as a set of self-consistent
stellar models, we shall deal with the problem of obtaining the corresponding
equilibrium DFs. By Jeans's theorem, they are functions of the isolating
integrals of motion that are conserved in each orbit. Some authors have shown
that, if the density can be written as a function of the gravitational
potential, it is possible to find such kind of two-integral DFs (see  Eddington
(1916); Fricke (1952); \cite{kal}; Jiang and Ossipkov (2007)). In this paper we
shall adopt the approach introduced by \cite{kal}, which fits quite well to
axisymmetric disc-like systems. Then, starting from the DFs derived from this
method, a second kind of DFs is obtained by using the formulae introduced by
\cite{dej}, that takes into account the principle of maximum entropy. These DFs
describe stellar systems with a preferred rotational state.

Accordingly, the paper is organized as follows. First, at section
\ref{sec:newfam}, we obtain the potential-density pairs for the new family of
thin disc models. Then, at section \ref{sec:kin}, we study the motion of test
particles around these new galactic models and, at section \ref{sec:DFs}, we
derive the distribution functions associated with the models. Finally, at
section \ref{sec:conc}, we summarize our main results.

\section{A New Family of Thin Disc Models}\label{sec:newfam}

\subsection{The generalized Kalnajs discs}\label{sec:kal}

In this subsection we summarize the principal features of the generalized
Kalnajs family, introduced by \cite{GR}, an infinite family of axially symmetric
finite thin discs. The mass surface density of each disc (labeled with the
positive integer $n$) is given by
\begin{equation}
\Sigma_{n}(R) = \Sigma_{c}^{(n)}\left[1 - \frac{R^{2}}{a^{2}}
\right]^{n-1/2}, \label{densidad}
\end{equation}
where $M$ is the total mass, $a$ is the disc radius and $\Sigma_{c}^{(n)}$ are
constants given by
\begin{equation}
\Sigma_{c}^{(n)}=\frac{(2n+1)M}{2\pi a^{2}}.
\end{equation}
Such mass distribution generates an axially symmetric gravitational potential,
that can be written as
\begin{equation}
\Phi_{n}(\xi,\eta)=-\sum_{j=0}^{m}C_{2j}q_{2j}(\xi)P_{2j}(\eta). \label{dk}
\end{equation}
Here, $P_{2j}(\eta)$ and $q_{2j}(\xi)=i^{2j+1}Q_{2j}(i\xi)$ are the usual
Legendre polynomials and the Legendre functions of the second kind respectively,
$-1 \leq \eta \leq 1$ and $0 \leq \xi < \infty$ are spheroidal oblate
coordinates, related to the usual cylindrical coordinates $(R,z)$ through the
relations
\begin{equation}
R^{2} = a^{2}(1 + \xi^{2})(1 - \eta^{2}), \qquad z = a\eta\xi, \label{tco}
\end{equation}
in such a way that the discs are located at $\xi = 0$ and $\eta =
\sqrt{1-R^{2}/a^{2}}$. Finally, the $C_{2j}$ are constants given by
$$
C_{2j}= \frac{M G\pi^{1/2} (4j+1) (2n+1)!}{a2^{2n+1} (2j+1) (n - j)!
\Gamma(n + j + \frac{3}{2} )q_{2j+1}(0)},
$$
where $G$ is the gravitational constant. 

As it was shown by \cite{GR}, in these models the surface density is a
monotonously decreasing function of the radius, with a maximum at the center of
the disk and vanishing at the edge, being the mass distribution of the  higher
members of the family more concentrated  at the center. On the other hand, the
corresponding rotation curves behave as follows: for $n=1$, the circular
velocity is proportional to the radius, whereas for the other members of the
family it increases from a value of zero at the center of the discs, then
reaches a maximum at a critical radius and, finally, decreases to a finite value
at the edge. Besides, the critical radius decreases as the value of $n$
increases.

\subsection{Formulation of the New Family}\label{sec:newfamfor}

Now, we will show that it is possible to formulate a new family of stellar
models by performing a linear combination of generalized Kalnajs discs, in such
a way that the new surface densities can be written as polynomials of the new
potentials. As it was quoted, this is a basic requirement for the derivation of
equilibrium DFs through the Kalnajs formalism sketched above. In particular, we
are interested on to derive a simple relation between the relative potential on
the disc and the surface density.

At first, note that $\Phi_n(0,\eta)$, given by (\ref{dk}), can be rewritten as
\begin{equation}
\Phi_n(0,\eta)=-\sum_{s=0}^{n}\sum_{r=0}^{s}A_{sr}\eta^{2s-2r},\label{eq:phi1}
\end{equation}
where $A_{sr}$ are constants defined as
\begin{equation}
A_{sr}=\frac{(-1)^r(4s-2r)!C_{2s}q_{2s}(0)}{2^{2s}r!(2s-2r)!(2s-r)!},
\end{equation}
and the relation (\ref{eq:phi1}) was derived by introducing the identity
(\cite{arf})
\begin{equation}
P_{2s}(\eta) = \sum_{r=0}^{s} \frac{(-1)^r(4s-2r)!}{2^{2s}r!(2s-2r)!(2s-r)!}
\eta^{2s-2r}. 
\end{equation}
From (\ref{eq:phi1}) we note that the maximum value of the gravitational
potential on the $n$th disc is $\Phi_n(0,0)$. Therefore, we define the relative
potential on such a disc  as
\begin{equation}
\Psi_n(\eta)=\Phi_n(0,0)-\Phi_n(0,\eta).\label{eq:psidef}
\end{equation}
Now, suppose that we can choose a linear combination of those $\Psi_n$ leading
to a new relative potential $\tilde{\Psi}_m$ of the form
\begin{equation}
\tilde{\Psi}_m=\sum_{n=1}^m B_n\Psi_n=A_{m0}\eta^{2m},\label{potnew}
\end{equation}
where $B_{n}$  are constants that can be determined from $A_{nr}$ and $C_{n}$
(see section \ref{sec:Bn}).

The new relative potential $\tilde{\Psi}_m$ is generated by a new mass
distribution described by a surface density $\tilde{\Sigma}_m$ that is also a
linear combination of generalized Kalnajs discs $\Sigma_{m}$. That is
\begin{equation}
\tilde{\Sigma}_m(R)=\sum_{n=1}^m
B_n\Sigma_c^{(n)}\eta^{2n-1}.\label{dennew}
\end{equation}
From this relation and (\ref{potnew}), we can note that
$\tilde{\Sigma}_m$ can be rewritten as
\begin{equation}
\tilde{\Sigma}_m(R)=\sum_{n=1}^m
B_n\Sigma_c^{(n)}\left(\frac{\tilde{\Psi}_m}{A_{m0}}\right)^{(2n-1)/(2m)}.\label{dennew2}
\end{equation}
Thus, we can see that this new family of discs is characterized by the fact that
the surface density can be split as a combination of powers of the relative
potential. This important fact makes viable the further derivation of two
integral DFs for the whole family (see section \ref{sec:DFs}), that can be
considered as a set of self-consistent galactic models. Now, the above
statements are only true if we can determinate the constants $B_{n}$, introduced
in (\ref{potnew}). In the next subsection, we show a procedure that, by using
the orthogonality properties of $P_{n}$, leads to a recurrence relation
expressing $B_{n}$ in terms of $A_{nr}$ and $C_{n}$.

\subsection{Calculation of  $B_n$}\label{sec:Bn}

According to definitions (\ref{eq:psidef}) and (\ref{dk}), the relative
potential associated to the generalized Kalnajs discs can be written as
\begin{equation}
\Psi_m(\eta)=\sum_{n=0}^{m}\tilde{C}_nP_{2n}(\eta),\label{eq:psileg}
\end{equation}
where $\tilde{C}_n$ are constants defined by
\begin{equation}
\tilde{C}_n=C_{2n}q_{2n}(0)-\delta_{0n}\sum_{i=0}^{m}C_{2i}q_{2i}(0)P_{2i}(0).
\end{equation}
So, by introducing (\ref{eq:psileg}) into (\ref{potnew}),
$\tilde{\Psi}_m$ can also be written in terms of Legendre
polynomials as
\begin{equation}
\tilde{\Psi}_m=\sum_{n=0}^m\sum_{i=0}^nB_n\tilde{C}_iP_{2i}(\eta)
=\sum_{n=0}^mD_nP_{2n}(\eta), \label{eq:psir}
\end{equation}
where $D_n$ are constants to be determined. Here, it is important to note that
these $D_n$ are such that
\begin{subequations}\begin{align}
D_m\,\;\;\, &=  B_m\tilde{C}_m,
\label{eq:4.22}   \\
D_{m-1} &= B_m\tilde{C}_{m-1}+B_{m-1}\tilde{C}_{m-1}, \label{eq:4.23}  \\
D_{m-2} &= B_m\tilde{C}_{m-2}+B_{m-1}\tilde{C}_{m-2}+B_{m-2}\tilde{C}_{m-2},\label{eq:4.24} \\
&\;\;\vdots\nonumber\\
D_{m-k} &= \tilde{C}_{m-k}\sum_{i=0}^{k}B_{m-i}.\label{eq:4.25}
\end{align}\end{subequations}
The above equations can be written in a compact way as
\begin{equation}
D_{n} = \tilde{C}_{n}\sum_{i=0}^{m-n}B_{m-i},\label{eq:4.25a}
\end{equation}
which summarizes the previous series of recurrence relations.

Now, let us come to the problem  of calculating $D_{n}$. According
to (\ref{potnew}) and (\ref{eq:psir}), we have
\begin{equation}
\sum_{n=0}^mD_nP_{2n}(\eta)=A_{m0}\eta^{2m},
\end{equation}
and by using the orthogonality properties of Legendre polynomials,
we obtain
$$
D_n=A_{m0}\frac{4n+1}{2}\int_{-1}^1\eta^{2m}P_{2n}(\eta)d\eta,
$$
that reduces to
\begin{equation}
D_n=A_{m0}\left[\frac{\pi^{1/2}(4n+1)\Gamma(2m+1)}{2^{2m+1}\Gamma(1+m-n)
\Gamma(m+n+\begin{matrix}\frac{3}{2}\end{matrix})}\right],
\end{equation}
in such a way that, by using (\ref{eq:4.25a}), we can determine the constants
$B_n$. In order to do this, note that equations (\ref{eq:4.22})-(\ref{eq:4.25})
give us recurrence relations: from (\ref{eq:4.22}) we obtain $B_m$, from
(\ref{eq:4.23}) we obtain $B_{m-1}$, and so on. In general, we have
\begin{subequations}\begin{align}
B_m\,\;\;\, &=  \frac{D_m}{\tilde{C}_m}=1,
\label{eq:422}   \\
B_{m-1} &= \frac{D_{m-1}}{\tilde{C}_{m-1}}-B_m, \label{eq:423}  \\
B_{m-2} &= \frac{D_{m-2}}{\tilde{C}_{m-2}}-B_m-B_{m-1},\label{eq:424} \\
&\;\;\vdots\nonumber\\
B_{m-k} &= \frac{D_{m-k}}{\tilde{C}_{m-k}}-\sum_{i=0}^{k-1}B_{m-i},
\quad\mathrm{for}\quad k\geq1,\label{eq:425}
\end{align}\end{subequations}
relations that can be summarized as
\begin{equation}
B_{n} = \frac{D_{n}}{\tilde{C}_{n}}-\sum_{i=0}^{m-n-1}B_{m-i},
\quad\mathrm{for}\quad n\leq m-1.\label{coefbn}
\end{equation}

\begin{figure}
\centering \epsfig{width=3in,file=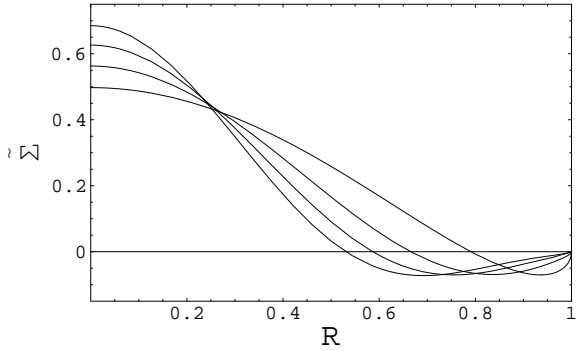}
\caption{We plot the mass surface density  $\tilde{\Sigma}_{2}$,
$\tilde{\Sigma}_{3}$, $\tilde{\Sigma}_{4}$ and $\tilde{\Sigma}_{5}$ (from bottom
to top in the left edge), given by (\ref{dennew2}), where the parameters $B_{n}$
are determined by (\ref{coefbn}). Since the resulting densities are negative in
some ranges, we must to correct $B_{1}$.}\label{fig:densneg}
\end{figure}

In table \ref{tabla1} we show the values of $B_n$ for the first four models,
i.e. $m=2,3,4,5$. Although we have solved the problem of finding these
constants, there is another inconvenient. If one introduces such coefficients 
in (\ref{dennew}), the corresponding surface density is negative for certain
ranges of $R$, as it is shown in figure \ref{fig:densneg}, where we plot the
resulting $\tilde{\Sigma}$ for the first four models. However, this problem can
be solved by correcting $B_{1}$, i.e. the coefficient corresponding to the
dominant term in (\ref{dennew}). In the next subsection, we show that it is
possible to define minimum values for $B_{1}$, corresponding to each model, in
such a way that all of they be described by positive surface densities.

\begin{table}
  \centering
  \begin{tabular}{|c|c|c|c|c|c|}
  \hline
  m & $B_{1}$ & $B_{2}$ & $B_{3}$ & $B_{4}$ & $B_{5}$\\
  \hline
  2& $-5/8$ & $1$ & & &\\
  3& $-35/192$ & $-7/12$ & $1$ & & \\
  4& $-105/1024$ & $-21/128$ & $-9/16$ & $1$ &  \\
  5& $-1155/16384$ & $-231/2560$ & $-99/640$ & $-11/20$ &  $1$ \\
  \hline
\end{tabular}
  \caption{Constants $B_{n}$ for the first four models: $m=2,3,4,5$} \label{tabla1}
\end{table}

\subsection{Correction of $B_1$}\label{sec:correctionB}

From here on, we consider $B_1$ as an arbitrary parameter that can be chosen in
such a way that $\tilde{\Sigma}\geq 0$, in the range $0\leq R\leq a$. For each
model, we expect that $B_1$ has a lower limit but does not has an upper limit.
The reason is that, according to (\ref{densidad}), we have
\begin{equation}
\lim_{R\rightarrow a}\frac{\mathrm{d}\Sigma_m}{\mathrm{d}R}=\left\{
        \begin{tabular}{cc}
        $-\infty$ & for $m=1$,\\
        $0$ & for $m\geq2$. \\
        \end{tabular}
\right.
\end{equation}
This means that the behavior of $\Sigma_{1}$, for $R\rightarrow a$, differs from
the remaining density surfaces, characterized by a rate of change tending
asymptotically to $0$ at the disc edge. Thus, it is evident that one always can
find a minimum value $B_{1\mathrm{min}}$, such that the product
$B_{1\mathrm{min}}\Sigma_{1}$ is larger than any linear combination of
$\Sigma_{2},\Sigma_{3},\ldots$.

In the particular case corresponding to the new models here formulated, the
surface density can be split as
\begin{eqnarray}
\tilde{\Sigma}_m(R,B_1)&=&B_1\Sigma_c^{(1)}\left(1-\frac{R^2}{a^2}\right)^{1/2}\nonumber\\
&&
+ \sum_{n=2}^m
B_n\Sigma_c^{(n)}\left(1-\frac{R^2}{a^2}\right)^{n-1/2},\label{dennew2}
\end{eqnarray}
where the coefficients $B_n$, for $n\geq2$, are given by (\ref{coefbn}). A
simple way to find $B_{1\mathrm{min}}$, is by demanding that $\tilde{\Sigma}_m$
has a minimum (equal to $0$) at $B_1=B_{1\mathrm{min}}$ and
$R=R_{\mathrm{min}}$. That is, we demand that the following two equations holds:
\begin{equation}
\left.\frac{\mathrm{d}\tilde{\Sigma}(R,B_{1\mathrm{min}})}
{\mathrm{d}R}\right|_{R=R_{\mathrm{min}}}=0,\label{sis1}
\end{equation}
\begin{equation}
\tilde{\Sigma}(R_{\mathrm{min}},B_{1\mathrm{min}})=0.\label{sis2}
\end{equation}
The relation (\ref{sis1}) imposes the condition that the surface density has a
minimum at $B_1=B_{1\mathrm{min}}$ and $R=R_{\mathrm{min}}$, while through the
relation (\ref{sis2}) we demand that its value at such critical point is $0$.

\begin{figure*}
\centering \epsfig{width=6.75in,file=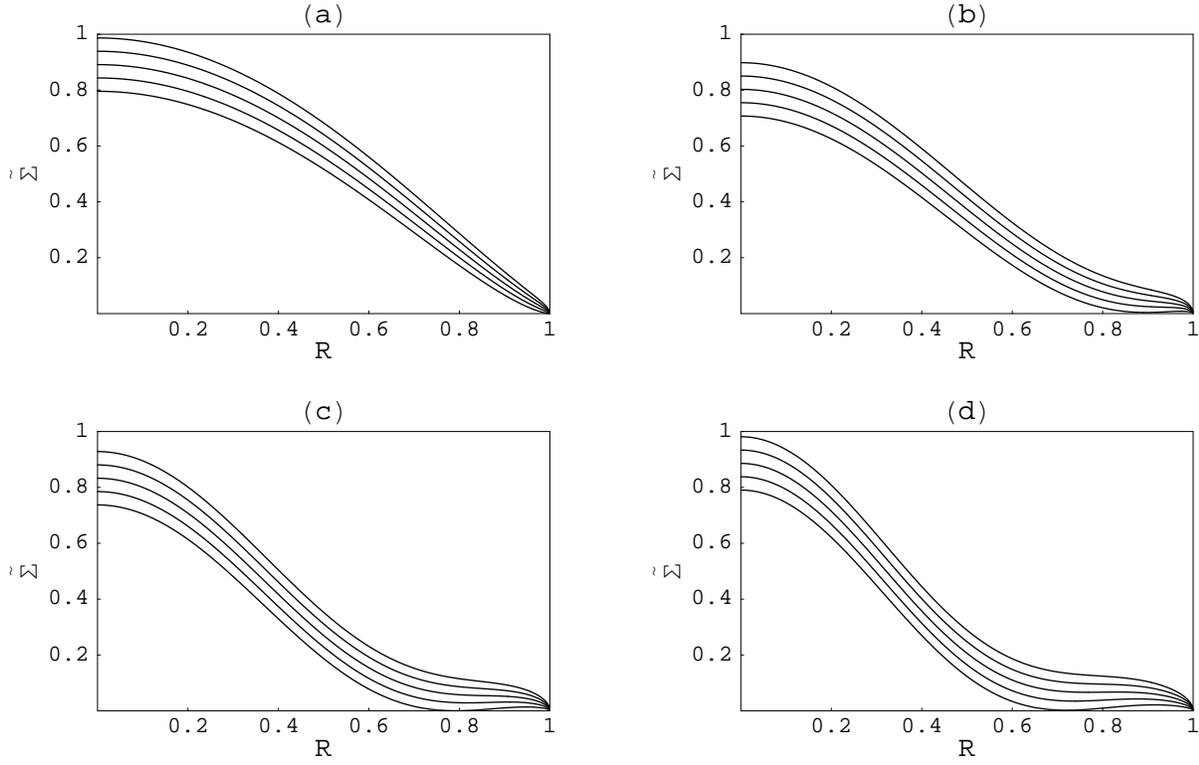}
\caption{We plot the mass surface density (a) $\tilde{\Sigma}_{2}$, (b)
$\tilde{\Sigma}_{3}$, (c) $\tilde{\Sigma}_{4}$, (d) $\tilde{\Sigma}_{5}$, for
different values of the parameter $B_{1}$. In each case we start with
$B_{1}=B_{1 min}$ (lower curve), given by table \ref{tabla2}. The remaining
curves, from bottom to top, corresponds to increments of $0.01$ in $B_{1}$,
starting from $B_{1 min}$.}\label{fig:Peddensi}
\end{figure*}

We find numerically the lower bounds of $B_{1}$ for the first four members of
the new family and they are showed in table \ref{tabla2}. The corresponding
surface densities, for different values of $B_1>B_{1\mathrm{min}}$, are plotted
in figure \ref{fig:Peddensi}. We note that, in all of these cases, the mass
concentration start in $0$ for $R=a$ and increases towards the disc bulge.
Moreover, such concentration increases with $m$.
\begin{table}
  \centering
  \begin{tabular}{|c|c|c|c|c|c|}
  \hline
  m & $B_{1 min}$ & $B_{2}$ & $B_{3}$ & $B_{4}$ & $B_{5}$\\
  \hline
  2& $0$ & $1$ & & &\\
  3& $0.101273$ & $-7/12$ & $1$ & & \\
  4& $0.128914$ & $-21/128$ & $-9/16$ & $1$ &  \\
  5& $0.143207$ & $-231/2560$ & $-99/640$ & $-11/20$ &  $1$ \\
  \hline
\end{tabular}
  \caption{Corrected constants $B_{n}$ for the first four models: $m=2,3,4,5$} \label{tabla2}
\end{table}
Finally, the gravitational potential corresponding to the new
family, derived as a linear combination of generalized Kalnajs
discs, can be cast as
\begin{equation}
\tilde{\Phi}_{m}(\xi,\eta)=-\sum_{n=1}^{m}\sum_{k=0}^{n}B_{n}C_{2k}q_{2k}(\xi)P_{2k}(\eta),\label{phinewgen}
\end{equation}
where $B_{n}$ are given by (\ref{coefbn}), for $n\geq 2$ and $B_{1}$
is an arbitrary parameter with lower bound determined by
(\ref{sis1})-(\ref{sis2}).

\section{Kinematics of the New Family}\label{sec:kin}

In this section we study the motion of test particles around the
galactic models formulated above. Since each $\tilde{\Phi}_{m}$ is
static and axially symmetric, the specific energy $E$ and the
specific axial angular momentum $\ell$ are conserved along the
particle motion. This fact restricts such motion to a three
dimensional subspace of the $(R,z,V_{R},V_{z})$ phase space. By
defining an effective potential $\Phi_{m}^{*}$ as
\begin{equation}
\Phi_{m}^{*} = \tilde{\Phi}_{m} +
\frac{\ell^{2}}{2R^{2}},\label{phieffect}
\end{equation}
the motion will be determined by the equations (\cite{BT})
\begin{subequations}\begin{align}
\dot{R} &= {V}_{R}, \label{em} \\
\dot{z} &= {V}_{z}, 	\\
\dot{V}_{R} &= -\frac{\partial \Phi_{m}^{*}}{\partial R}, \\
\dot{V}_{z} &= -\frac{\partial \Phi_{m}^{*}}{\partial z}, \label{em4}
\end{align}\end{subequations}
together with
\begin{equation}
E = \frac{1}{2}({V}_{R}^{2}+{V}_{z}^{2}) +
\Phi_{m}^{*},\label{totalenergy}
\end{equation}
which gives the total energy of the particle.

Relations (\ref{em})-(\ref{totalenergy}) are the basic equations
that determine the motion of a particle with specific axial angular
momentum $\ell$ and energy $E$, in cylindrical coordinates. At
first, we restrict our attention on particles belonging the disc,
i.e. the interior kinematics, in order to describe rotation curves
(for circular orbits) and the stability of nearly circular orbits
(by deriving the epicyclic and vertical frequencies). Then, we shall
focus on three dimensional motion, in particular, the case of disc
crossing orbits. In such case, we use $z=0$ surfaces of section, in
order to illustrate the regularity or chaoticity characterizing
those orbits.

\subsection{Interior Kinematics}\label{subsec:kin1}

\begin{figure*}
\centering \epsfig{width=6.75in,file=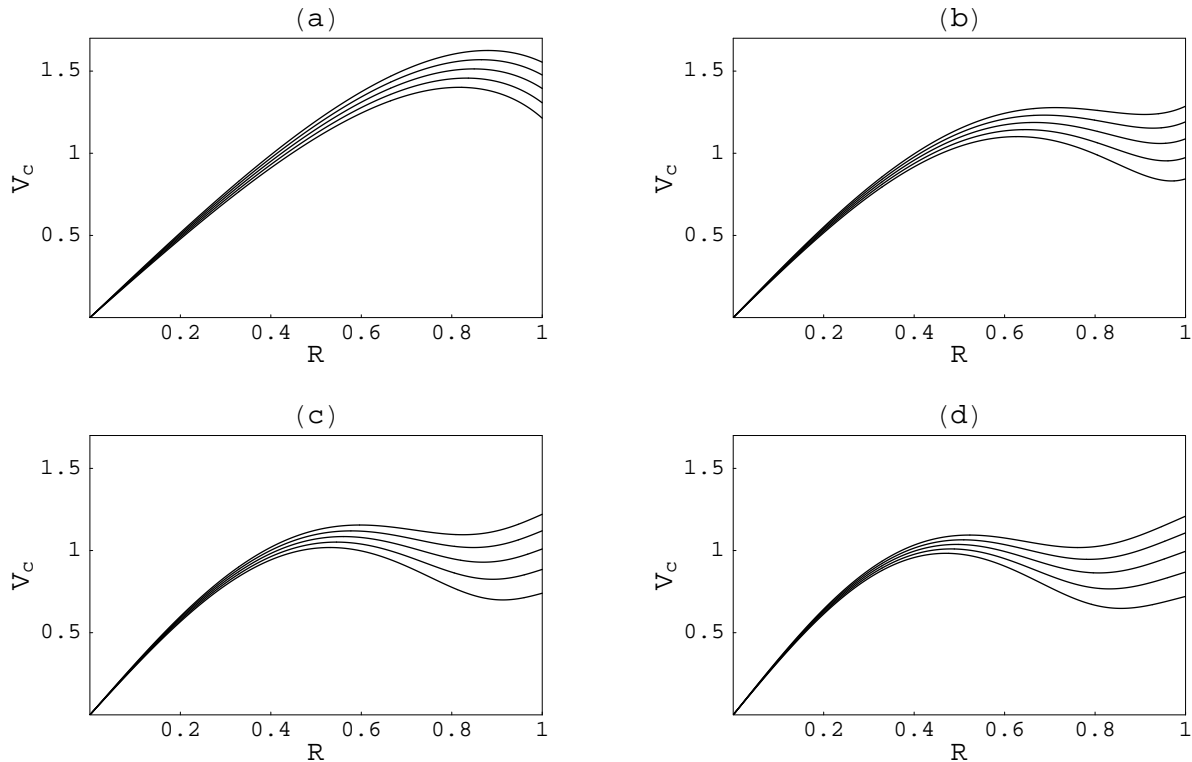}
\caption{We plot the rotation curves for (a) $m=2$, (b) $m=3$, (c) $m=4$, (d)
$m=5$, for the same values of parameter $B_{1}$ as in figure \ref{fig:Peddensi}.
The lower curves corresponds to $B_{1 min}$ and the upper curves corresponds to
larger values of $B_{1}$.} \label{fig:PedVc}
\end{figure*}

We start by pointing out that the system (\ref{em}) has equilibrium
points at $V_{R}=V_{z}=z=0$, $R=R_{c}$, where $R_{c}$ must satisfy
the equation
\begin{equation}
\left(\frac{\partial \Phi_{m}^{*}}{\partial R}\right)_{(R_{c},0)}=
-\frac{\ell^{2}}{R_{c}^{3}}+\left(\frac{\partial
\tilde{\Phi}_{m}}{\partial R}\right)_{(R_{c},0)}=0,\label{circular}
\end{equation}
that is the condition for a circular orbit in the plane $z=0$. In
other words, the equilibrium points of (\ref{em}) occur when the
test particle describes equatorial circular orbits of radius
$R_{c}$, specific axial angular momentum given by
\begin{equation}
\ell_{c}=\pm \sqrt{R_{c}^{3}\left(\frac{\partial
\tilde{\Phi}_{m}}{\partial R}\right)_{(R_{c},0)}}\:\:,\label{Lz}
\end{equation}
and specific energy
\begin{equation}
E=\Phi_{m}^{*}(R_{c},0).\label{Ecircular}
\end{equation}
The subscript $c$ in $\ell_{c}$ indicates that we are dealing with
circular orbits.

\begin{figure*}
\centering \epsfig{width=6.75in,file=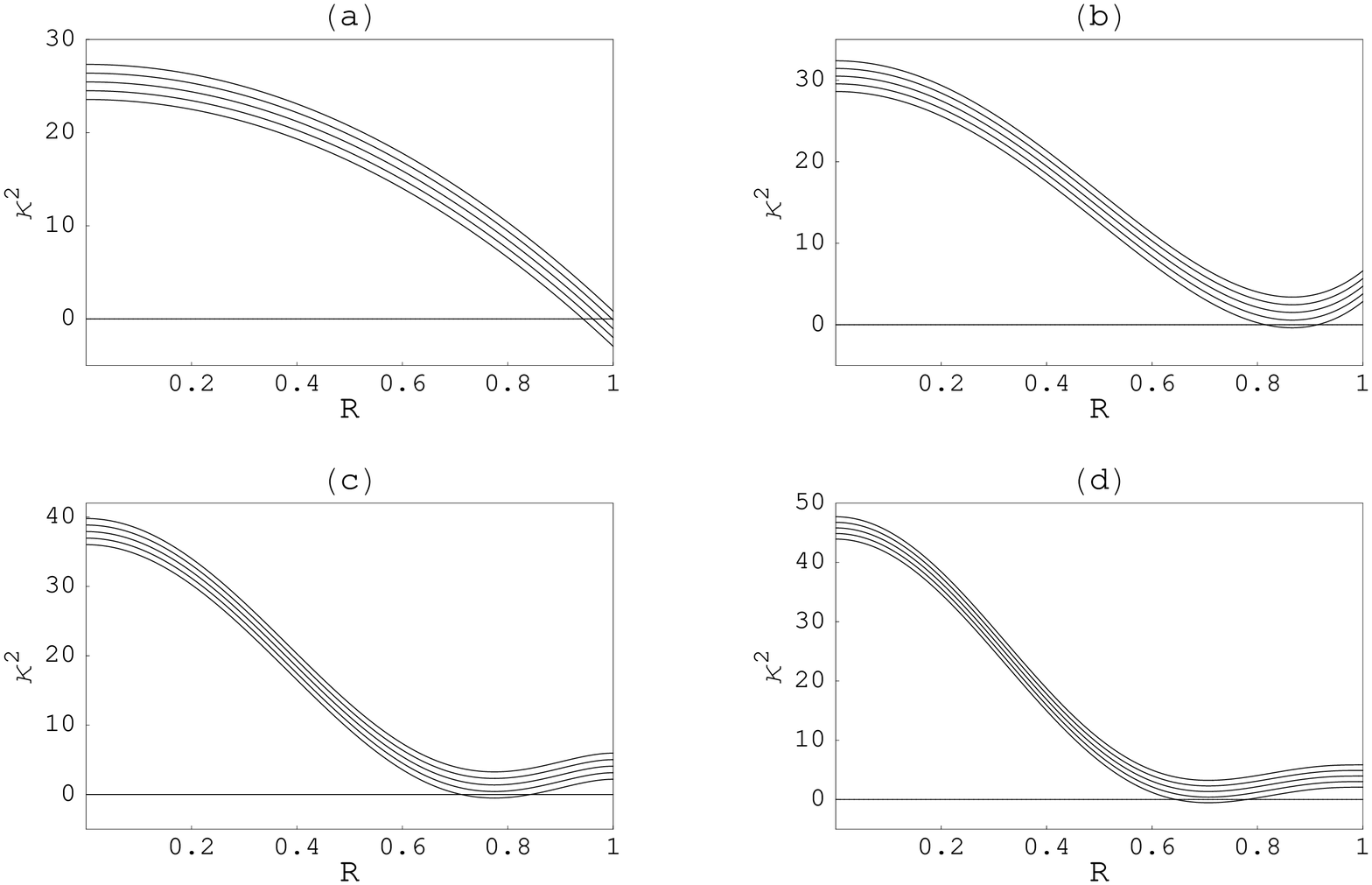}
\caption{Quadratic epicyclic frequency as a function of $R$ for the cases (a)
$m=2$, (b) $m=3$, (c) $m=4$, (d) $m=5$. We use the same values of parameter
$B_{1}$ as in figure \ref{fig:Peddensi}. The lower curves corresponds to $B_{1
min}$ and the upper curves corresponds to larger values of $B_{1}$. In all of
these cases, we note a prominent range of radial stability. Such range increases
with $B_{1}$.}\label{fig:PedEpi}
\end{figure*}
\begin{figure*}
\centering \epsfig{width=6.75in,file=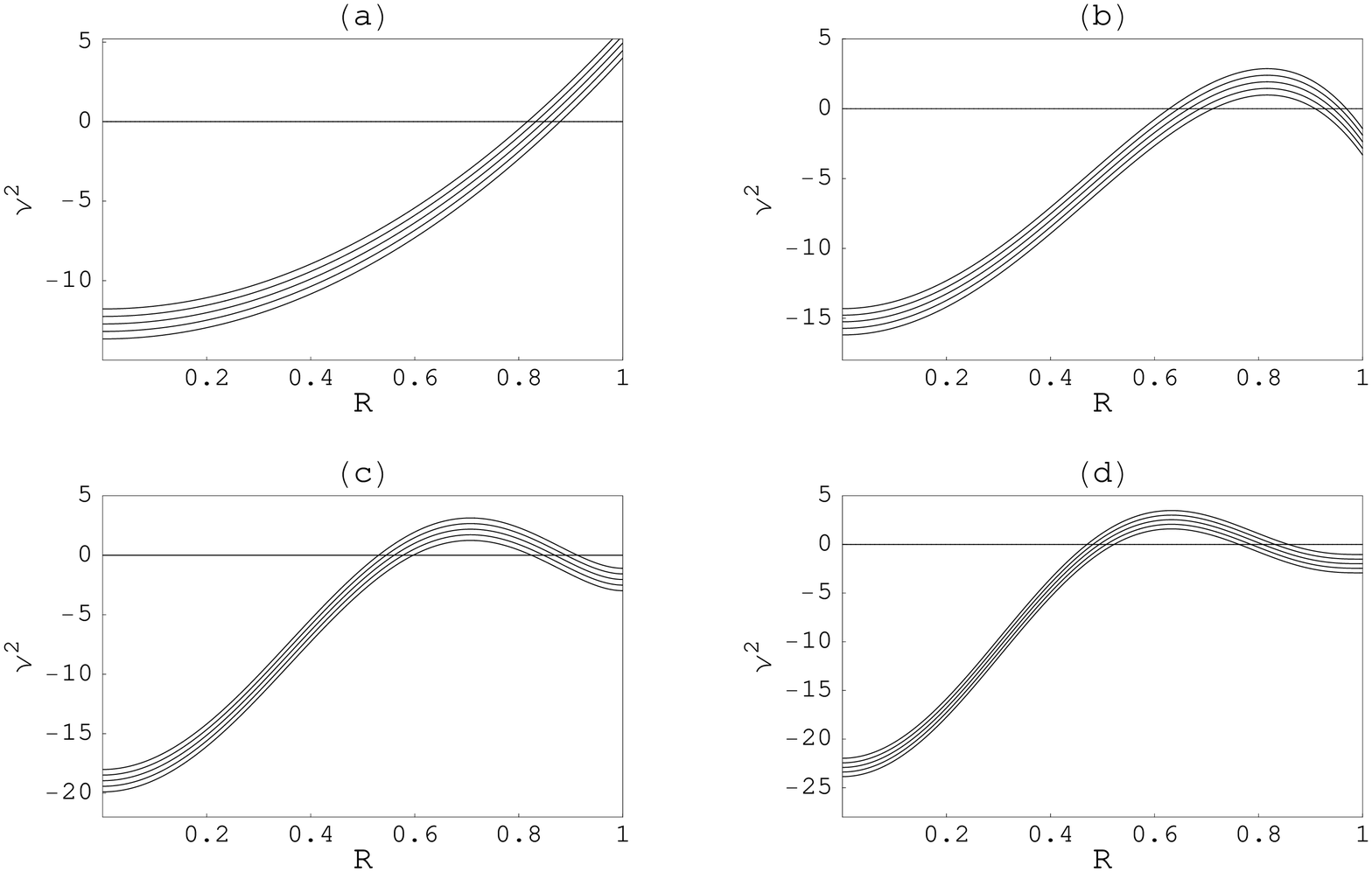}
\caption{Quadratic vertical frequency as a function of $R$ for the cases (a)
$m=2$, (b) $m=3$, (c) $m=4$, (d) $m=5$. We use the same values of parameter
$B_{1}$ as in figure \ref{fig:Peddensi}. The lower curves corresponds to $B_{1
min}$ and the upper curves corresponds to larger values of $B_{1}$. The range of
vertical stability is small and increases with $B_{1}$.}\label{fig:PedVert}
\end{figure*}

Now, a feature of special interest is the circular velocity $V_{c}=\ell_{c}
R_{c}$ (sometimes denoted by $V_{\varphi}$), which can be directly compared with
astronomical observations. Then, it is convenient to express $V_{c}$ as a
function of the radius. From (\ref{Lz}), we can write the magnitude of the
circular velocity as $V_{c}^{2}=R\left(\partial \tilde{\Phi}_{m}/\partial
R\right)_{z=0}$. In figure \ref{fig:PedVc}, we plot $V_{c}(R)$ corresponding to
the first four models of the new family of discs. For $B_1=B_{1\mathrm{min}}$
the rotation curves has a maximum and then decreases to a constant value, in a
similar fashion that the behavior of generalized Kalnajs discs (\cite{GR}).
Moreover, the maximum of the rotation curve is at an even smaller radius when
the value of m increases. Now, in the case in which $B_1$ is very large, the
curve approximates to a straight line, as a consequence of the dominance of term
associated with the usual Kalnajs disc. On the other hand, for intermediate
values of $B_1$ we note an interesting feature: when the rotation curve reaches
its maximum, it remains nearly constant, as it is the case in many real
galaxies.

In order to study the stability of these trajectories under small radial and
vertical ($z$-direction) perturbations, we analyze the nature of quasi-circular
orbits. They are characterized by an epicycle frequency $\kappa$ and a vertical
frequency $\nu$, given by (Binney \& Tremaine \citeyear{BT})
\begin{equation}
\kappa^{2}=\left(\frac{\partial^{2} \Phi_{m}^{*}}{\partial
R^{2}}\right)_{(R_{c},0)}, \:\:\:\:\:\:\:\:\:\:
\nu^{2}=\left(\frac{\partial^{2} \Phi_{m}^{*}}{\partial
z^{2}}\right)_{(R_{c},0)}.\label{epicicle}
\end{equation}
This means that by introducing (\ref{Lz}) in the second derivatives of
$\Phi_{m}^{*}$ we obtain $\kappa^{2}$ and $\nu^{2}$ as functions of the radius.
Values of $R$ such that $\kappa^{2}>0$ and (or) $\nu^{2}>0$ corresponds to
stable circular orbits under small radial and (or) vertical perturbations,
respectively. In figures \ref{fig:PedEpi} and \ref{fig:PedVert} we show the
behavior of the epicycle and vertical frequency, respectively, for the first
four models and using the same values of $B_{1}$ as in figure
\ref{fig:Peddensi}. We note that these models are characterized by a prominent
range of stability under radial perturbations. In particular, for $B_{1}>B_{1
min}$, there will be radially stable orbits with radius in the range $0\leq
R\leq a$. In contrast, there are prominent ranges of vertical instability. Such
ranges tend to decreases when $m$ and $B_{1}$ increases. Thus, we can say that
the stability under vertical perturbations, in quasi-circular orbits, improves
in models with a large $m$.

\subsection{Exterior Kinematics: Disc-crossing Orbits}\label{subsec:kin2}

In this section, we study the behavior of 3-dimensional motion, i.e.
orbits outside the equatorial plane (except when they cross the
plane $z=0$). As it was mentioned above, the motion is determined by
(\ref{em}) and can be described in an effective phase space with
three dimensions (there are two integrals of motion: $E$ and
$\ell$). An adequate tool to investigate such orbits is the
Poincar\'e surface of section, in order to find the chaotic and (or)
regular regions characterizing the structure of the phase space.

In particular, we present numerical solutions of (\ref{em})- (\ref{em4}) for the
case of bounded disc-crossing orbits. There are certain values of $E$ and $\ell$
for which they are confined to regions that contain the disc and will cross back
and forth through it. As it was showed by Hunter (\citeyear{Hunter}), this fact
usually gives rise to many chaotic orbits due to the discontinuity in the
$z$-component of the gravitational field, producing a fairly abrupt change in
their curvatures. There is an important exceptional case of this behavior: the
Kuzmin's disc, characterized by an integrable potential of the form
$\Phi=-GM[R^{2}+(a+|z|)^{2}]^{-1/2}$, with $a>0$. However, the so-called
Kuzmin-like potentials, characterized by $\Phi(\varepsilon)$ where
$\varepsilon=[R^{2}+(a+|z|)^{2}]^{1/2}$, are non-integrable and present the
behavior mentioned above. The family of models formulated here present a very
similar structure and we can expect an analogous dynamics. Each potential
$\tilde{\Phi}_{m}(\xi,\eta)$ can be cast in a Kuzmin-like form if we take into
account that, according to (\ref{tco}), $\xi=(R_{+}+R_{-})/2a$ and
$\eta=(R_{+}-R_{-})/2ia$, where $R_{+}=[R^{2}+(z+ia)^{2}]^{1/2}$ and
$R_{-}=[R^{2}+(z-ia)^{2}]^{1/2}$. Moreover, they are characterized by a
z-derivative discontinuity in the disc, given by
\begin{equation}
\left(\frac{\partial\tilde{\Phi}_{m}}{\partial
z}\right)_{z=0^{+}}=-\left(\frac{\partial\tilde{\Phi}_{m}}{\partial
z}\right)_{z=0^{-}}=2\pi G\:\: \tilde{\Sigma}_{m}
(R).\label{discont}
\end{equation}
Despite the above relation makes the KAM theorem inapplicable, we also found a
large variety of regular disc-crossing orbits.

\begin{figure}
\centering \epsfig{width=2.5in,file=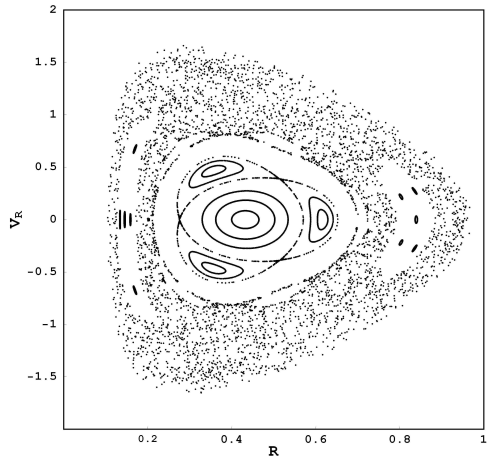}
\caption{Surface of section for some orbits with $\ell=0.2$, $E=-1.245$, around
the model $m=2$ with $B_{1}=0.1$.}\label{fig:PedOrbit1}
\end{figure}

\begin{figure}
\centering \epsfig{width=2.5in,file=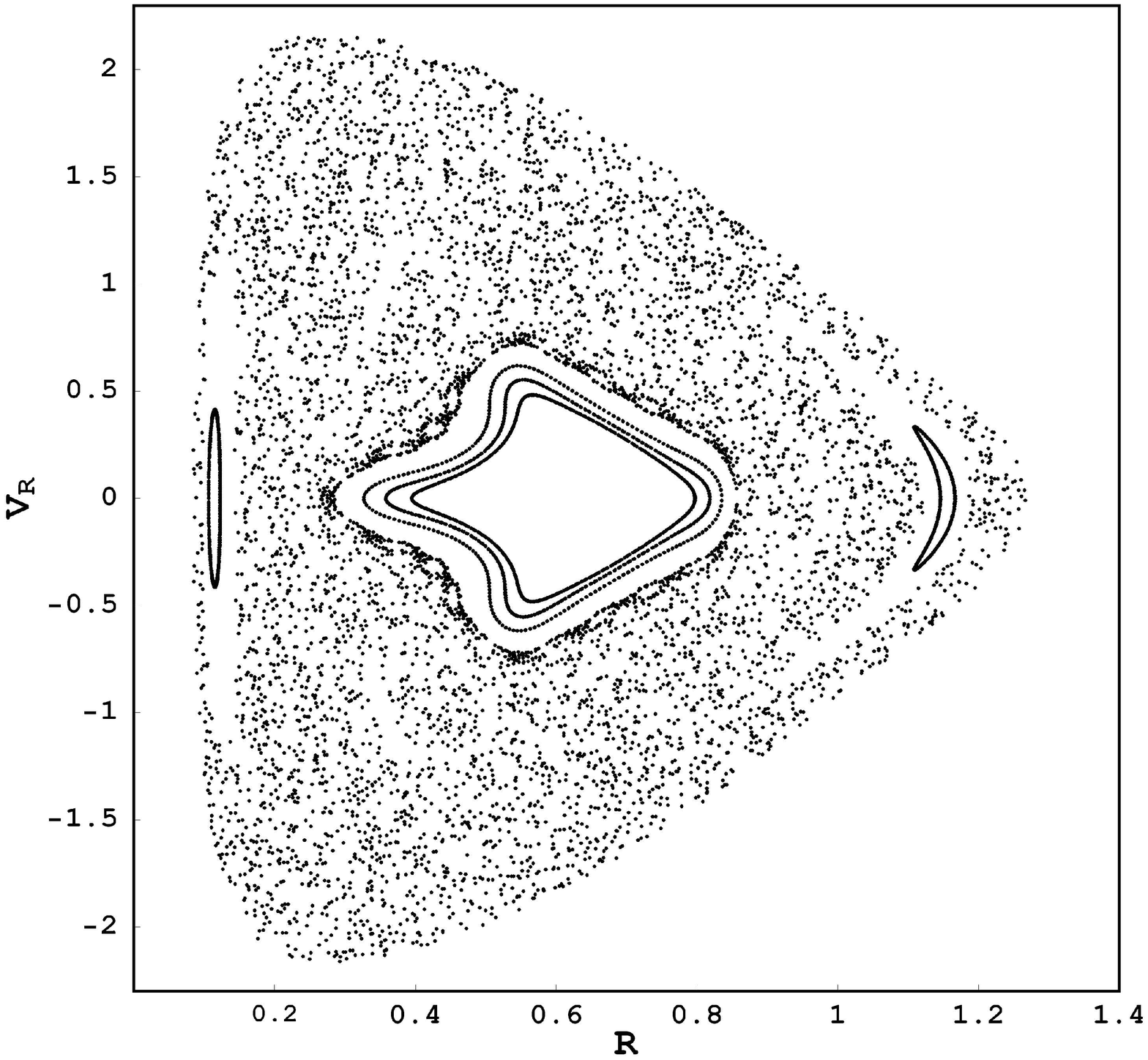}
\caption{Surface of section for some orbits with $\ell=0.2$, $E=-1.245$, around
the model $m=2$ with $B_{1}=0.5$.}\label{fig:PedOrbit2}
\end{figure}

In Figure \ref{fig:PedOrbit1}, we show the $z=0$-surface of section
corresponding to some orbits with $E=-1.245$ and $\ell=0.2$ (these same values
are used to perform the next three surfaces of section), corresponding to a test
particle moving around the model $m=2$ with $B_{1}=0.1$. As it was expected,
from our experience with the generalized Kalnajs discs (\cite{ramos}), this plot
exhibits a variety of regular and chaotic trajectories. There is a regular
central region conformed by two kinds of KAM curves: the central rings made by
box orbits; a set of resonant islands  chain (made by loop orbits) enclosing the
rings. Moreover, there are two lateral regular zones of loop orbits that, as
well as the central region, are enclosed by a sea of chaos. In figure
\ref{fig:PedOrbit2}, we show the effect of increasing $B_{1}$ in model $m=2$.
Then, for the same initial conditions with $E=-1.245$ and $\ell=0.2$, the
surface of section reveals an increasing in the chaotic region along with a
distortion of the KAM curves in regular zones (for example, now the central
torus are made only by box orbits).

\begin{figure}
\centering \epsfig{width=2.5in,file=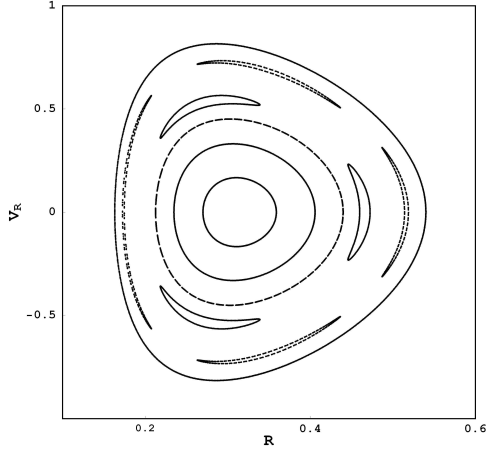}
\caption{Surface of section for some orbits with $\ell=0.2$, $E=-1.245$, around
the model $m=3$ with $B_{1}=0.2$. In this case we have only regular
orbits}\label{fig:PedOrbit3}
\end{figure}

\begin{figure}
\centering \epsfig{width=2.5in,file=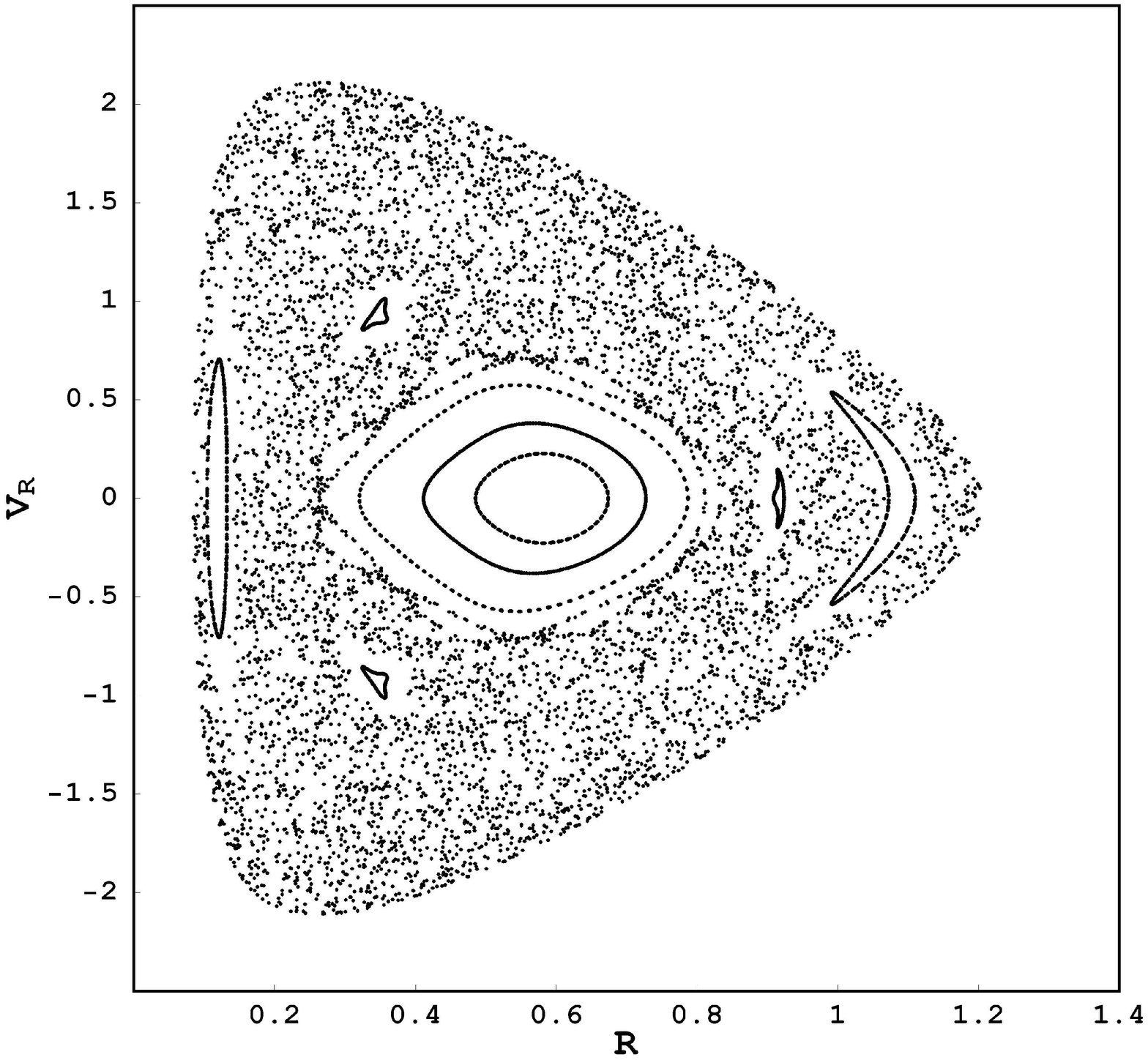}
\caption{Surface of section for some orbits with $\ell=0.2$, $E=-1.245$, around
the model $m=3$ with $B_{1}=1$. Now, a prominent chaotic region appears as a
consequence of the overlapping of the lateral island chains, showed in the
previous figure.}\label{fig:PedOrbit4}
\end{figure}

One could expect similar chaotic surfaces of section for models $m=3,\ldots$,
but what really happens is that for certain values of $B_{1}$ all the orbits are
regular. This is the case illustrated in figure \ref{fig:PedOrbit3}, where the
Poincar\'e section, corresponding to a particle moving around the model $m=3$,
reveals completely regular motion. This is a surprising fact, since in this kind
of potentials chaos is the rule. If one increases $B_{1}$ (for example, from 0.2
to 1), as in the case of figure \ref{fig:PedOrbit4}, we find again a prominent
chaotic region enclosing some regular regions of KAM curves. Thus, the free
parameter $B_{1}$, apart from determining some relevant features in equatorial
orbits, plays an essential role in 3-dimensional motion and we can conjecture
that, for certain values, it is possible to have a third integral of motion.


\section{Two-Integral Distribution Functions}\label{sec:DFs}

The derivation of the DFs, associated to the models constructed above, is
particularly simple if we work in an adequate rotating frame, as it was pointed
out by \cite{kal}. To begin with, we had seen that for a $B_1$ given by the
recurrence relation (\ref{coefbn}) (from here on we denote it as $B_1^{*}$) the
relative potential would be
\begin{equation}
\tilde{\Psi}_m^{*}=B_1^{*}\Psi_1+\sum_{n=2}^mB_n\Psi_n=A_{m0}\eta^{2m}.
\end{equation}
On the other hand, if we choose a different $B_1$, the relative potential
becomes
\begin{equation}
\begin{split}
\tilde{\Psi}_m &= B_1\Psi_1 + \sum_{n=2}^mB_n\Psi_n = A_{m0}\eta^{2m} +
(B_1-B_1^{*})\Psi_1,\\
&= A_{m0}\eta^{2m}+\begin{matrix} \frac{1}{2}
\end{matrix}(B_1-B_1^{*})\Omega_{0}^{2}a^2\eta^2,
\end{split}
\end{equation}
where $\Omega_0=[3\pi GM/(4a^3)]^{1/2}$.

Now, let's work in a a rotating frame with angular velocity $\Omega$. In this
frame, the effective potential is defined as (\cite{BT})
\begin{equation}
\Phi_e=\Phi-\begin{matrix} \frac{1}{2}
\end{matrix}\Omega^2R^2=\Phi+\begin{matrix} \frac{1}{2}
\end{matrix}\Omega^2a^2(\eta^2-1),
\end{equation}
and by choosing properly the constants, the relative-effective potential will be
\begin{equation}
\tilde{\Psi}_{e,m}=A_{m0}\eta^{2m}+\begin{matrix} \frac{1}{2}
\end{matrix}(B_1-B_1^{*})\Omega_{0}^{2}a^2\eta^2-\begin{matrix} \frac{1}{2}
\end{matrix}\Omega^2a^2\eta^2.\label{prelef}
\end{equation}
Note that if one chooses $\Omega$ as
\begin{equation}
\Omega=\pm\Omega_0\sqrt{B_1-B_1^{*}},\label{omega}
\end{equation}
the term with $\eta^2$ vanishes and equation (\ref{prelef}) reduces to
\begin{equation}
\tilde{\Psi}_{e,m}=A_{m0}\eta^{2m}.
\end{equation}
Therefore, the relation between the density and the relative-effective potential
can be written as
\begin{equation}
\tilde{\Sigma}_m(R)=\sum_{n=1}^mB_n\Sigma_c^{(n)}
\left(\frac{\tilde{\Psi}_{e,m}}{A_{m0}}\right)^{(2n-1)/(2m)}.
\end{equation}
Finally, by using the Kalnajs method (\cite{kal}), we obtain that the DF
corresponding to the $m$-model is given by
\begin{equation}
f_{m}(\varepsilon,L_z)=\frac{1}{4\pi
m}\sum_{n=1}^m\frac{B_n\Sigma_c^{(n)}(2n-1)A_{m0}^{-(2n-1)/(2m)}}
{\left(\varepsilon+\Omega L_{z}-\begin{matrix} \frac{1}{2}
\end{matrix}\Omega^{2}a^{2}\right)^{1-(2n-1)/(2m)}}.\label{dfsnew2}
\end{equation}

The explicit DFs for the first 4 models and the associated $\Omega$, given by
(\ref{omega}), are
\begin{subequations}\begin{align}
f_2&=\frac{3\Sigma_c^{(2)}A_{20}^{-3/4}}{4J^{1/4}}+\frac{B_1
\Sigma_c^{(1)}A_{20}^{-1/4}}{4J^{3/4}},\label{newDF2} \\&\nonumber\\
f_3&=\frac{5\Sigma_c^{(3)}A_{30}^{-5/6}}{6J^{1/6}}-\frac{7
\Sigma_c^{(2)}A_{30}^{-3/6}}{24J^{3/6}}+\frac{B_1
\Sigma_c^{(1)}A_{30}^{-1/6}}{6J^{5/6}},\label{newDF3} \\&\nonumber\\
f_4&=\frac{7\Sigma_c^{(4)}A_{40}^{-7/8}}{8J^{1/8}}-\frac{45\Sigma_c^{(3)}A_{40}^{-5/8}}{128J^{3/8}}\nonumber\\&\nonumber\\
&-\frac{63
\Sigma_c^{(2)}A_{40}^{-3/8}}{1024J^{5/8}}+\frac{B_1\Sigma_c^{(1)}A_{40}^{-1/8}}{8J^{7/8}},\label{newDF4} \\&\nonumber\\
f_5&=\frac{9\Sigma_c^{(5)}A_{50}^{-9/10}}{10J^{1/10}}-\frac{77\Sigma_c^{(4)}A_{50}^{-7/10}}{200J^{3/10}}
\nonumber\\&\nonumber\\&-\frac{99\Sigma_c^{(3)}A_{50}^{-5/10}}{1280J^{5/10}}-\frac{693\Sigma_c^{(2)}A_{50}^{-3/10}}{25600J^{7/10}}+\frac{B_1
\Sigma_c^{(1)}A_{50}^{-1/10}}{10J^{9/10}},\label{newDF5}
\end{align}\end{subequations}
where $J=\varepsilon+\Omega L_{z}-\begin{matrix} \frac{1}{2}
\end{matrix}\Omega^{2}a^{2}$ is the Jacobi's integral and
\begin{subequations}\begin{align}
A_{20} &=\frac{45 G M \pi }{128 a},&\Omega_{2}^{2}&=\frac{3 \pi
G M (5 +8 B_1)}{32 a^3},\label{beta} \\&&&\nonumber\\
A_{30} &=\frac{175 G M \pi }{512 a},&\Omega_{3}^{2}&=\frac{3 \pi  G M(35 +192 B_1)}{768 a^3},\label{gamma} \\&&&\nonumber\\
A_{40}  &=\frac{11025 G M \pi }{32768 a},&\Omega_{4}^{2}&=\frac{3 \pi  G M(105+1024 B_1)}{4096 a^3},\label{mu} \\&&&\nonumber\\
A_{50} &=\frac{43659 G M \pi }{131072 a},&\Omega_{5}^{2}&=\frac{3
\pi G M(1155+16384 B_1)}{65536 a^3}.\label{nu}
\end{align}\end{subequations}

\begin{figure*}
\centering \epsfig{width=7in,file=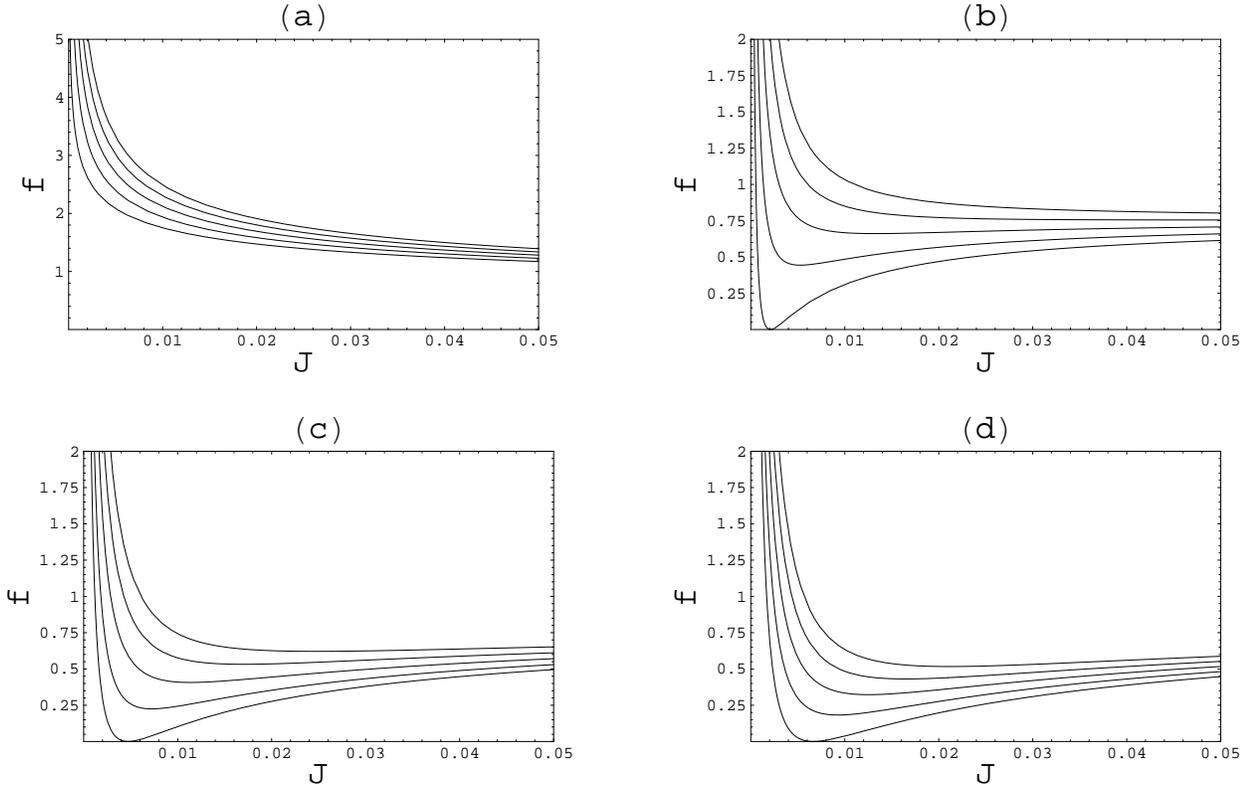} \caption{We plot the
DFs as functions of the Jacobi's integral for (a) $m=2$, (b) $m=3$, (c)
$m=4$, (d) $m=5$, for different values of the
parameter $B_1$. The lower curves corresponds to $B'_{1min}$ and the upper curves corresponds to larger values of $B_1$}\label{fig:DFJAC}
\end{figure*}

It is easy to see that the DFs obtained by equation (\ref{dfsnew2}), for the
cases $m\geq3$, could be negative in some regions of the phase space
corresponding to the physical domain. To avoid this inconvenient, it is
necessary to impose a stronger condition for the constants $B_1$, in order to
obtain  well-defined distribution functions. To do this, we  formulate the
following equations (in a similar fashion as in subsection
\ref{sec:correctionB}):
\begin{equation}
\left.\frac{\mathrm{d}f_m(J,B_{1\mathrm{min}}')}
{\mathrm{d}R}\right|_{J=J_{\mathrm{min}}}=0,\label{sist1}
\end{equation}
\begin{equation}
f_m(J_{\mathrm{min}},B_{1\mathrm{min}}')=0.\label{sist2}
\end{equation}
Relation (\ref{sist1}) imposes the condition that the DF has a minimum at
$B_1=B'_{1\mathrm{min}}$ and $R=R_{\mathrm{min}}$, while through the relation
(\ref{sist2}), we demand that its value at such critical point vanishes. The
numeric solution to these equations give us the values shown in table
\ref{tablabp}, for the models with $m=3,4,5$.
\begin{table}
  \centering
  \begin{tabular}{|c|c|}
  \hline
  m & $B'_{1 min}$ \\
  \hline
  2& $0$ \\
  3& $0.182292$ \\
  4& $0.287827$ \\
  5& $0.381061$ \\
  \hline
\end{tabular}
  \caption{Coefficients $B'_{1 min}$ for the first four models: $m=2,3,4,5$} \label{tablabp}
\end{table}
So, taking this values as a lower limit for $B_1$, the DFs given by
(\ref{dfsnew2}) become positive-defined in the physical domain of the phase
space. The figure \ref{fig:DFJAC} shows the graphics of the DFs as functions of
the Jacobi's integral, with different values of $B_1$. In general, we can
observe that the probability is maximum for small values of $J$, and tends to a
constant as $J$ increases. Moreover, in the cases $m\geq3$ we can see that for
values of $B_1$ near to $B'_{1\mathrm{min}}$, the probability has a minimum at
$J\approx J_{\mathrm{min}}$, and it is zero for $B_1=B'_{1\mathrm{min}}$ at $J=
J_{\mathrm{min}}$, in agreement with equations (\ref{sist1}) and (\ref{sist2}).

\begin{figure*}
\centering \epsfig{width=6.75in,file=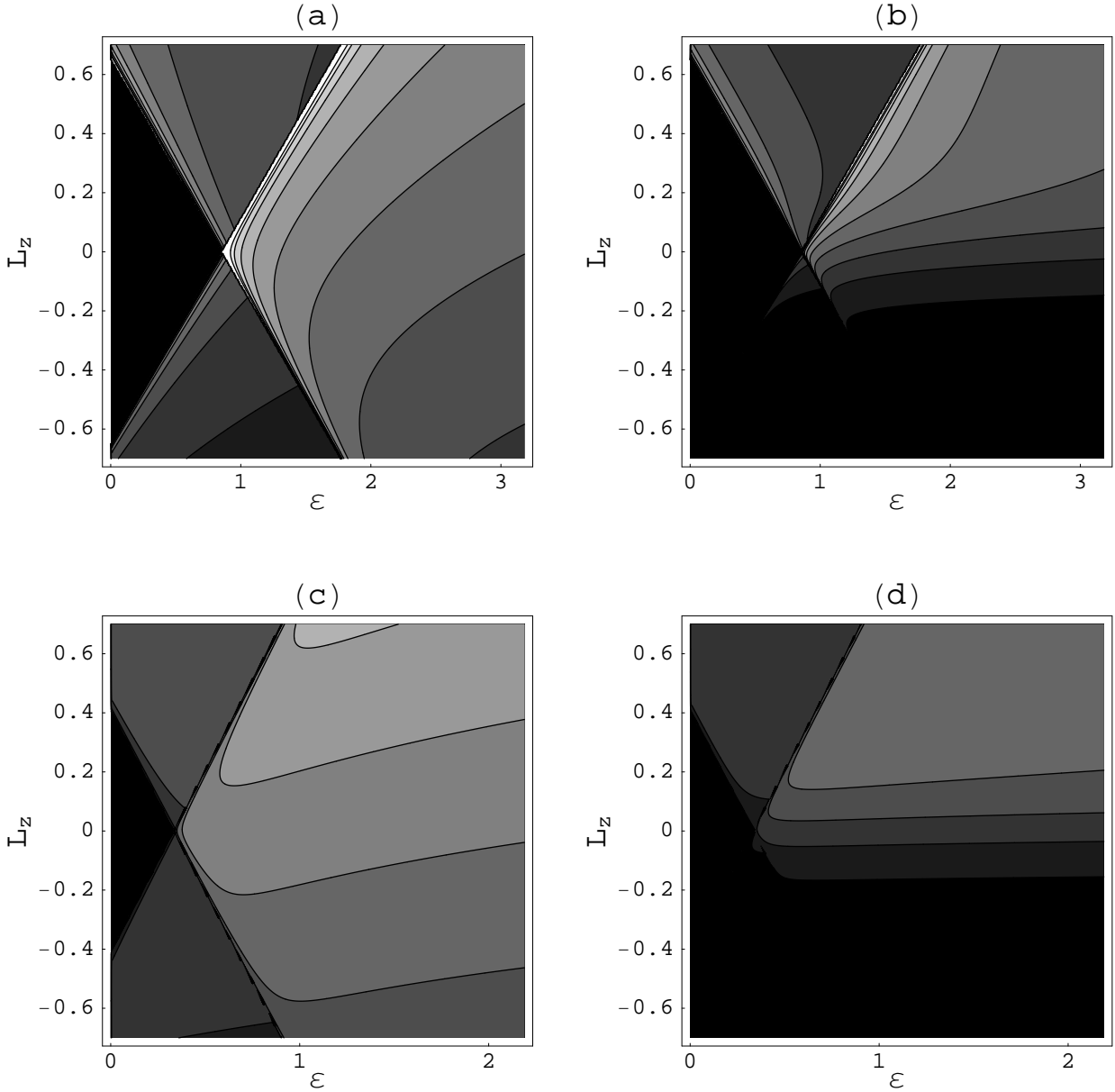} \caption{Contours of
$\tilde{f}_{m}$, given by (\ref{dfsnew3}), for $m=2$, $B_1=0.1$ and (a) $\alpha=1$; (b)
$\alpha=10$. Moreover, it is plotted the case $m=3$ with $B_1=0.2$ and (c) $\alpha=1$; (d) $\alpha=10$. Larger values of the
DF corresponds to lighter zones.}\label{fig:PedDF1}
\end{figure*}

On the other hand, it is convenient to derive a new kind of DFs, corresponding
to more probable rotational states. As it was shown by \cite{dej}, it is
possible to obtain DFs obeying the maximum entropy principle, through the
equation
\begin{equation}
\tilde{f}_{m}(\varepsilon,L_z)=\frac{2f_{m+}(\varepsilon,L_{z})}{1+e^{-\alpha
L_{z}}},\label{dfsnew3}
\end{equation}
where $f_{m+}$ is the even part of (\ref{dfsnew2}). The figure \ref{fig:PedDF1}
shows the behavior of the DFs given by (\ref{dfsnew3}). In (a) and (b) are
plotted the contours corresponding to the model $m=2$, with different values of
parameter $\alpha$. As it can be seen, $\alpha$ determines a particular
rotational state in the stellar system. As $\alpha$ increases, the probability
to find a star with positive $L_{z}$ increases as well. A similar result can be
obtained for $\alpha<0$, when the probability to find a star with negative
$L_{z}$ decreases as $\alpha$ decreases, and the corresponding plots would be
analogous to figure \ref{fig:PedDF1}, after a reflection about $L_{z}=0$. In (c)
and (d) are plotted the contours of model $m=3$ with $B_1\approx
B'_{1\mathrm{min}}$ for the same values of $\alpha$. The behavior of the DFs for
the remaining cases is pretty similar to the shown in these figures: when
$B_1\gg B'_{1\mathrm{min}}$, the contours are similar to (a) and (b), while if
$B_1\approx B'_{1\mathrm{min}}$, the contours are similar to (c) and (d), in
agreement with figure \ref{fig:DFJAC}.

\section{Concluding Remarks}\label{sec:conc}

We have obtained a set of models for axisymmetric flat galaxies, by superposing
members belonging to the generalized Kalnajs discs family. The mass distribution
of each model (labeled through the parameter $m=2,3,\ldots$), described by
(\ref{dennew}), is maximum at the center and vanishes at the edge, in
concordance with a great variety of galaxies. Moreover, the mass density can be
expressed as a function of the gravitational potential (see equation
(\ref{dennew2})), which makes possible to derive, analytically, the equilibrium
DFs describing the statistical features of the models.

These models have also interesting features concerning with the interior
kinematical behavior. On one hand, we showed that for some values of $B_{1}$,
the circular velocity has a behavior very similar to that seen in many discoidal
galaxies. This is a very relevant fact, which suggests that it is not always 
necessary to introduce the hypothesis of dark matter halos (or MOND theories) in
order to describe adequately a variety of rotational curves. On the other hand,
the analysis of epicyclic and vertical frequencies, associated to quasi-circular
orbits, reveals that the models are stable under radial perturbations but
unstable under vertical disturbances. With regard to the motion of test
particles around the models formulated here, we found that the behavior of
disc-crossing orbits is similar to that seen in the generalized Kalnajs family.
However, for certain values of the parameter $B_{1}$, the Poincar\'e surface of
section reveals that one can suggest the existence of a (non analytical) third
integral of motion.

On the other hand, we find two kinds of equilibrium DFs for the models. Such
two-integral DFs can be formulated, at first, as functionals of the Jacobi's
integral, as it was sketched in the formalism developed by \cite{kal}. This
class of DFs essentially describes systems which rotational state, in average,
behaves as a rigid body. Then, we use the procedure introduced by \cite{dej},
obtaining DFs which represents systems with a mean rotational state consistent
with the maximum entropy principle and, therefore, more probable than the first
ones. The statements exposed above suggest that the family presented here, can
be considered as a set of realistic models that describes satisfactorily a great
variety of galaxies.

\section{Acknowledgments}

J. R-C. thanks the financial support from {\it Vicerrector\'ia
Acad\'emica}, Universidad Industrial de Santander.


\begin{thebibliography}{9999}


\bibitem[\protect\citeauthoryear{Arfken}{2005}]{arf} Arfken G. \& Weber H., 2005,  Mathematical Methods for
Physicists. 6th Ed., Academic Press.

\bibitem[\protect\citeauthoryear{Bagin}{1987}]{bagin} Bagin V. M., 1972,  Astron. Zhur., 49, 1249.

\bibitem[\protect\citeauthoryear{Binney \& Tremaine}{1987}]{BT} Binney, J. and
Tremaine, S., 1987,  Galactic Dynamics. Princeton University Press,
Princeton, N. J.

\bibitem[\protect\citeauthoryear{Brandt \& Belton}{1962}]{BB} Brandt, J. C.
and Belton, M. J. S., 1962, Ap. J., 136, 352

\bibitem[\protect\citeauthoryear{Dejonghe}{1986}]{dej} Dejonghe, H., 1986, Phys. Rep., 133 (3\& 4),
217.

\bibitem[\protect\citeauthoryear{Evans}{1993}]{eva93} Evans N. W., 1993, MNRAS, 260, 191.

\bibitem[\protect\citeauthoryear{Evans}{1994}]{eva94} Evans N. W., 1994,  MNRAS, 267, 333.

\bibitem[\protect\citeauthoryear{Fricke}{1962}]{fricke} Fricke, W.
, 1952, Astron. Nachr., 280, 193-216.

\bibitem[\protect\citeauthoryear{Gonz\'alez \& Reina}{2006}]{GR} Gonz\'alez, G. A. and Reina, J. I. 2006, MNRAS, 371 (4), 1873-1876.

\bibitem[\protect\citeauthoryear{Hunter}{1963}]{HUN1} Hunter, C., 1963, MNRAS,
126, 299.


\bibitem[\protect\citeauthoryear{Hunter \& Quian}{1993}]{Hunter} Hunter C. \& Quian E. 1993,
MNRAS, 262 , 401-428.

\bibitem[\protect\citeauthoryear{Jiang}{2000}]{jiang00} Jiang Z., 2000, MNRAS, 319, 1067.

\bibitem[\protect\citeauthoryear{Jiang \& Moss}{2002}]{jiangmos02} Jiang Z. \& Moss D., 2002, MNRAS, 331, 117.

\bibitem[\protect\citeauthoryear{Jiang \& Ossipkov}{2006}]{jiangoss06} Jiang Z. \& Ossipkov L., 2006, Astron.
Astroph. Trans., 25, 213.

\bibitem[\protect\citeauthoryear{Jiang \& Ossipkov}{2007}]{jiang} Jiang Z. \& Ossipkov L., 2007, MNRAS,
379 (3),1133-1142.

\bibitem[\protect\citeauthoryear{Kalnajs}{1972}]{KAL} Kalnajs, A. J., 1972,
Ap. J., 175, 63.

\bibitem[\protect\citeauthoryear{Kalnajs}{1976}]{kal} Kalnajs, A. J., 1976, Ap. J., 205,
751.

\bibitem[\protect\citeauthoryear{Kutuzov}{1995}]{Kut1} Kutuzov S. A., 1995, Astron. Astroph. Trans., 7, 191.

\bibitem[\protect\citeauthoryear{Kutuzov}{1980}]{Kut2} Kutuzov S. A., Ossipkov L. P., 1980,  Astron. Zhur., 57, 28.

\bibitem[\protect\citeauthoryear{Kutuzov}{1986}]{Kut3} Kutuzov S. A., Ossipkov L. P., 1986, Translated from Astrofizika,
25(3), 545-558.

\bibitem[\protect\citeauthoryear{Kutuzov}{1988}]{Kut4} Kutuzov S. A., Ossipkov L. P., 1988,  Astron. Zhur., 65, 468.

\bibitem[\protect\citeauthoryear{Kuzmin}{1956}]{KUZ} Kuzmin, G., 1956, Astron.
Zh., 33, 27.

\bibitem[\protect\citeauthoryear{Linden-Bell}{1962}]{LB} Linden-Bell, D., 1962, MNRAS,
123, 447-458.


\bibitem[\protect\citeauthoryear{Miyamoto}{1971}]{Miya} Miyamoto M., 1971, Publ. Astron. Soc.  Japan, 23, 21.

\bibitem[\protect\citeauthoryear{Miyamoto \& Nagai}{1975}]{Miyamoto} Miyamoto M. \& Nagai R., 1975, Publ. Astron. Soc.  Japan,  27, 533.

\bibitem[\protect\citeauthoryear{Miyamoto \& Nagai}{1976}]{Nagai} Nagai R. \& Miyamoto M., 1976, Publ. Astron. Soc.  Japan,  28, 1.


\bibitem[\protect\citeauthoryear{Pedraza, Ramos-Caro \& Gonz\'alez}{2008}]{PRG}
Pedraza J. F., Ramos-Caro J. \& Gonz\'alez G. A., 2008, submitted to MNRAS.

\bibitem[\protect\citeauthoryear{Ramos-Caro, L\'opez-Suspez \&
Gonz\'alez}{2008}]{ramos} Ramos-Caro J., L\'opez-Suspez F. \& Gonz\'alez G. A., 
2008, MNRAS, 386, 440-446.

\bibitem[\protect\citeauthoryear{Schmith}{1956}]{Schmith} Schmith, M., 1956,
Bull. Astron. Inst. Neth, 13, 15.

\bibitem[\protect\citeauthoryear{Toomre}{1963}]{T1} Toomre, A., 1963, Ap. J.,
138, 385.

\bibitem[\protect\citeauthoryear{Toomre}{1964}]{T2} Toomre, A., 1964, Ap. J.,
139, 1217

\bibitem[\protect\citeauthoryear{Wyse \& Mayall}{1942}]{WM} Wyse, A. B. and
Mayall, N. U., 1942, Ap. J., 95, 24









\end{thebibliography}
\end{document}